\documentclass[12pt,a4paper]{article}
\usepackage[textwidth=450pt, textheight=660pt]{geometry}
\usepackage{amsmath,amsthm,amssymb}
\usepackage{graphicx,tabu}
\usepackage[pagebackref=false]{hyperref}
\usepackage[active]{srcltx}
\usepackage[affil-it]{authblk}
\usepackage{wrapfig}
\usepackage[noadjust]{cite}
\usepackage{filecontents}
 \allowdisplaybreaks
 \newcommand{\be}{\begin{equation}}
\newcommand{\ee}{\end{equation}}
\newcommand{\bea}{\begin{eqnarray}}
\newcommand{\eea}{\end{eqnarray}}
\newcommand{\nn}{\nonumber}

\newcommand{\trm}{\textrm}
\newcommand{\tit}{\textit}

\newcommand{\ba}{\begin{array}}
\newcommand{\ea}{\end{array}}
\newcommand{\bfig}{\begin{figure}}
\newcommand{\efig}{\end{figure}}

\newcommand{\hs}{\hspace}
\newcommand{\bi}{\begin{itemize}}
\newcommand{\ei}{\end{itemize}}
\newcommand{\bc}{\begin{cases}}
\newcommand{\ec}{\end{cases}}
\newcommand{\bal}{\begin{aligned}}
\newcommand{\eal}{\end{aligned}}
\newcommand{\ben}{\begin{enumerate}}
 \newcommand{\een}{\end{enumerate}}

\begin{document}

  \vspace{2cm}

  \begin{center}
    \font\titlerm=cmr10 scaled\magstep4
    \font\titlei=cmmi10 scaled\magstep4
    \font\titleis=cmmi7 scaled\magstep4
  {\bf On  the high temperature limit of the Casimir energy }

    \vspace{1.5cm}
     \noindent{{\large Y. Koohsarian  \footnote{yo.koohsarian@mail.um.ac.ir}, K. Javidan  \footnote{Javidan@um.ac.ir}
  }} \\
      {\it Department of Physics, Ferdowsi University of Mashhad \\
       P.O.Box 1436, Mashhad, Iran} \\

  \end{center}
  \vskip 2em

\begin{abstract}
We introduce a useful approach   to find    asymptotically explicit expressions for the Casimir free energy  at large temperature. The resulting expressions  contain the classical terms as well as the few first terms of the corresponding heat-kernel expansion, as expected. This technique works well for many familiar configurations in Euclidean as well as non-Euclidean spaces.  By utilizing this approach, we provide some new numerically considerable results   for the Casimir pressure in some rectangular ideal-metal cavities.   For instance, we show  that at sufficiently large temperature, the Casimir pressure acting on the sidewalls of  a rectangular  tube   can be up to twice that of the two parallel  planes. We also apply this technique for calculating the Casimir  free energy  on a $3$-torus as well  as a $3$-sphere. We show that a nonzero mass term for both scalar and  spinor fields as well on the torus  as on the sphere, violates the third law of thermodynamics.  We obtain some  negative values for the Casimir entropy   on the $3$-torus as well as on the $3$-sphere.   We speculate  that these negative Casimir entropies can be interpreted thermodynamically as an instability of the vacuum state at finite temperatures. 
 \end{abstract}


\section{Introduction} \label{sec-1}
It is well known that the total  zero-point  energy of  a quantum  field  restricted by some   physical   boundaries, can have some macroscopic manifestations, such as the famous Casimir effect \cite{Cas}.  Nowadays various aspects of the  Casimir effect   have  been   investigated extensively   for various configurations, see \cite{CR1,CR2,CR3,CR4,CR5,CR6} as reviews.  One of the important,   still controversial  \cite{CT1,CT2,CT3,CT4,CT5}  aspects   of the Casimir  energy   is its temperature  dependence.   In the framework  of the quantum field theory, as we know, the vacuum energy at finite temperature is obtained generally by applying the known Matsubara imaginary-time  formalism. This formalism is actually  equivalent   to the quantum thermodynamics approach, where the finite-temperature Casimir energy, known also as the Casimir free energy, is given by the total free energy of the quantized field modes, see e.g. \cite{CR1,CR2,CR3,CR4}. 

 The representations obtained in the literature for the   Casimir  free energy  usually have no asymptotically  explicit  expressions at high temperatures. Conventionally  the high-temperature asymptotic expression of the Casimir energy  is obtained by applying the known  \tit{heat kernel} expansion,  which has  some complications e.g. in calculating the classical terms \cite{CR3}. In this paper, we introduce another useful approach to obtain   exact representations for the Casimir energy having  asymptotically explicit  expression for the high-temperature limits. This is a useful technique specially for fully bounded configurations, such as   the rectangular cavities, which are familiar  configurations in the Casimir effect  literature  \cite{CRec1,CRec2,CRec3,CRec4,CRec5,CRec6,CRec7,CRec8}, also for some compact  manifolds with nontrivial topology and curvature such as torus and sphere, which    have significant role in some cosmological models \cite{CR1,CR3}.  In this approach all the high-temperature limiting terms including the classical terms as well as the main first  terms of the heat-kernel expansion would be automatically obtained in exact  forms.  These terms would have explicit dependence on the mass parameter, providing asymptotically suitable expressions as well for small as for large masses. Such asymptotically explicit expressions might not be simply obtained through the conventional  approaches, such as the heat-kernel expansion.\\
 
  In section 2,  we introduce a rather general approach  to find   representations asymptotically suitable     at high temperatures,  for the Casimir free energy.   We use the mentioned approach to obtain new representations  for the Casimir  free energy,  for  the  scalar     as well as the electromagnetic  field  in configurations with flat boundary conditions specifically in rectangular cavities (section 3).  We   provide some new     numerical results for the Casimir pressures in rectangular tube and rectangular box, and show that at sufficiently large temperature, the Casimir pressure in the tube and the box can be larger, by factors of $2$ and $1.5$ respectively, than that of the parallel planes. In section 4, we apply the mentioned approach for the scalar as well as the spinor field in the  non-Euclidean space, specifically on a $3$-torus and a $3$-sphere. We show that a nonzero mass term for both scalar and  spinor fields as well on the torus  as on the sphere, violates the third law of thermodynamics.  In some cases we obtain negative values for the Casimir entropies, and interpret these negative entropies as an instability in the vacuum state.

\section{ A   useful approach  } \label{sec-2}

  From the functional formalism of the quantum field theory, the  vacuum  energy of a scalar  field restricted by some appropriate boundary conditions, can be written as  \cite{CR3}
 \be
E_0= -\frac{i }{2}  \int_{-\infty}^{\infty} \frac{d\xi}{2\pi}  \sum_J  \ln\left[-\xi^2+ k_J^2+\mu^2 \right]
\label{e_0}
 \ee
where,  ``$\omega$'' and ``$k$''  represent the energy and the momentum of    off-mass-shell  scalar modes, and  the collective index $J$ labels the mode numbers.  At finite temperature,   the mode frequencies, $\omega$'s,   are replaced with the (imaginary)  Matsubara frequencies 
\be
  -i\omega_l=\frac{2\pi}{\beta} l \ \ \ ; \ \ \ l=0,\pm1,\pm2,... 
  \label{Mf}
\ee
 in which  $\beta\equiv 1/T$, with $T$ as the temperature. As a a result, the scalar  free  energy  \eqref{e_0}  takes the form \cite{CR1,CR2,CR3,CR4}
 \be
E_0(T) = \frac{ 1}{2 \beta} \sum_J \sum_{l=-\infty}^{\infty}   \ln\left[\left(\frac{2\pi}{\beta} \right)^2 l^2+ k_J^2+\mu^2  \right]. 
\label{e_0(T)}
 \ee
 which after regularization and subtracting the contribution of the free vacuum, gives the Casimir free energy of the scalar field. For  a Dirac 4-spinor,  in topologically flat spaces,  one should just multiply  the above equation by ``$-4$''.
 
  Now, as is conventional, by introducing a  regularization parameter, the scalar free energy \eqref{e_0(T)}  takes the form
 \bea
 E_0 =  -\frac{1}{2\beta}  \lim_{s\to 0} \frac{\partial}{\partial s}  \sum_{l=-\infty}^{\infty}   \sum_{J}  \left[  l^2+ \lambda_J ^2+\lambda_\mu^2  \right]^{-s}  
 \label{pe_0(T)}
\eea
 where by discarding an irrelevant constant term, we have introduced  the dimensionless parameters $\lambda_J\equiv  \beta k_J/2\pi $ and  $\lambda_\mu \equiv \beta\mu/2\pi$.  A  sum of the form   \eqref{pe_0(T)},  can be written in terms of  an inhomogeneous  Epstein-like zeta function \cite{math5} as 
 \bea
&& Z_{p,\{b_1,b_2,...,b_p\}} \left(a_1 , a_2, ...,a_p, c; s\right) \equiv \sum_{\{n_i\}=0  }  \left[ \sum_{i=1}^p a_i^2 (n_i+b_i)^2 +c^2\right]^{-s}; \nn \\
 &&\hs{8cm}  s<0, \ \ a_1\leq a_2\leq...\leq a_p.
 \label{z_p}
 \eea
 Then using the familiar gamma function, as is conventional, we rewrite the above expression as a parametric integral;
  \bea
  Z_{p,\{b_i\}} \left(a_i ,c; s\right)= \sum_{\{n_i\}=0}^\infty  \int_0^\infty \frac{dt}{t} \frac{t^s}{\Gamma(s)} 
 \exp \left[-t\left( \sum_{i=1}^p a_i^2 (n_i+b_i)^2 +c^2\right)\right]
  \label{pz_p}
  \eea
Subsequently   we use  a generalized form of the Poisson summation formula  \cite{math1},
 \bea
&&\sum_{n=0}^{\infty} \exp\left[-\alpha^2\left(n+\theta\right)^2 \right]=\sum_{n=0}^{\infty} \frac{(-1)^n}{n!}\alpha^{2n} \zeta (-2n,\theta)  +  \frac{\sqrt{\pi}}{2\alpha}\nn \\ 
&& \hspace{7cm}+ \frac{\sqrt{\pi}}{\alpha}\cos(2\pi \theta)\sum_{n=1}^{\infty} \exp\left[-\frac{n^2\pi^2}{\alpha^2}\right].
\label{sf}
\eea
 in which, 
 \be
\zeta (s,r) =\sum_{n=0}^{\infty} (n+r)^{-s}
\label{Hz}
\ee
is the known Hurwitz zeta function.  Now we apply the formula \eqref{sf} just to the $n_i$-sum with the smallest $a_i$. 
 This condition, as we see in the next sections, is necessary for obtaining asymptotically suitable expressions for the Casimir energy  as well  for low as for high temperature limits.  So here by applying Eq. \eqref{sf} to the $n_1$-sum in Eq. \eqref{pz_p} one can find
 \bea
&& Z_{p,\{b_{1,2,...,p}\}} \left(a_1,  a_2 ,...,a_p ,c; s\right) =\nn \\
&&\hs{1cm} \sum_{n_1=0}^{\infty} \frac{(-1)^{n_1}}{n_1!}a_1^{2n_1} \zeta (-2n_1,b_1) \frac{\Gamma(s+n_1)}{\Gamma(s)}Z_{p-1,\{b_{2,3,...,p}\}}\left(   a_2 ,...,a_p ,c; s\right) \nn \\
&&  +\frac{\sqrt{\pi}}{2a_1 \Gamma(s)} \Big( \Gamma\left(s-1/2\right) Z_{p-1,\{b_{2,3,...,p}\}} (a_2,b_2,...,a_p,b_p,c;s-1/2)   \nn \\
&&\hs{6cm}  +  G_{p,\{b_{2,3,...,p}\}}(a_1,a_2, ...,a_p , c;s-1/2)\Big)
 \label{z_p-re}
 \eea
 in which 
\bea
&& G_{p,\{b_{2,3,...,p}\}}(a_1,a_2 ,...,a_p ,c;s)\equiv 4 \sum_{ n_1=1}^\infty \ \sum_{n_2,...,n_p=0}^\infty   \left(\frac{\sum_{i=2}^p a_i^2 (n_i+b_i)^2+c^2}{\pi^2   n_1^2 /a_1^2}\right)^{-s/2}\nn \\
&&\hs{5cm} \times K_{-s}\left[2\pi \left(\left[\sum_{i=2}^p a_i^2 (n_i+b_i)^2+c^2\right]  n_1^2 /a_1^2\right)^{1/2}\right]
\label{G}
\eea
with
\be
K_\nu(z) = \frac{(z/2)^\nu \Gamma(1/2)}{\Gamma(\nu+1/2)} \int_1^\infty e^{-zt} \left(t^2-1 \right)^{(2\nu-1)/2} dt,
\label{Bf}
\ee
 as a  modified Bessel function of the second kind, and we have used the following integral relation
  \be
\int_0^{\infty} \frac{dt}{t} t^{-r} \exp\left[-x^2 t-y^2/t \right]dt=2(x/y)^{r}K_{r}(2xy).
  \label{ir}
\ee
 Now as is seen Eq. \eqref{z_p-re} has a recurrence form, see Eq. (4.19) of \cite{math2},  hence we need just to find a regularized form for $Z_1$. But
  such a  sum can be directly regularized by utilizing   the known   Abel-Plana formula \cite{math5}, so  by using the recurrence \eqref{z_p-re} one can find exact regularized expressions for other $Z$ functions. 
  
  In this work we are interested specifically in the cases with $b_i=0,\frac12$. In theses cases by  using
  \bea
 \zeta (0,0)=\frac12, \  \ \ \zeta (-2n,0)=0\ ; \ \ \  n=1,2,... \nn \\
 \zeta (-2n,1/2)=0 \ ; \ \  n=0,1,..., \nn
 \eea
 the recurrence \eqref{z_p-re}  can be simplified as
 \bea
&& Z_{p,0} \left(a_1,  a_2 ,...,a_p  ,c; s\right) =-\frac12 Z_{p-1,0}(a_2,a_3, ... , a_p, c, s)\nn \\
 &&  \hs{1cm} +\frac{\sqrt{\pi}}{2a_1 \Gamma(s)} \Big( \Gamma(s-1/2) Z_{p-1,0}  (a_2, ...,a_p, c; s-1/2)   \nn \\
&&\hs{6cm}  +  G_{p,0} (a_1,a_2, ...,a_p,  c;s-1/2)\Big)
 \label{rz_p0-re} \\
&& Z_{p,\frac12} \left(a_1,   a_2, ...,a_p,c; s\right) = \nn \\
 &&  \hs{1cm} \frac{\sqrt{\pi}}{2a_1 \Gamma(s)} \Big( \Gamma(s-1/2) Z_{p-1,\frac12} (a_2, ...,a_p, c;s-1/2)   \nn \\
&&\hs{6cm}  - G_{p,\frac12}(a_1,a_2, ...,a_p,  c;s-1/2)\Big)
 \label{rz_p1/2-re}
 \eea 
 in which 
 \bea
&& Z_{p,0} \left(a_1, a_2, ...,a_p,  c; s\right) \equiv \sum_{\{n_i\}=1  }  \left[ \sum_{i=1}^p a_i^2 n_i ^2 +c^2\right]^{-s} \label{z_p0} \\
&&  Z_{p,\frac12} \left(a_1, a_2, ...,a_p,  c; s\right) \equiv \sum_{\{n_i\}=0  }  \left[ \sum_{i=1}^p a_i^2 (n_i+1/2) ^2 +c^2\right]^{-s}
\label{z_p1/2}
 \eea
and $G_{p,x} \equiv G_{p,\{b_{2,3,...,p}=x\}}$,  see Eq. \eqref{G}.    Now the  recurrences \eqref{rz_p0-re} and  \eqref{rz_p1/2-re} start with the two simplest cases
 \bea
&& Z_{1,0}(a_1, c;s)=\sum_{n_1=1}^\infty\left[a_1^2 n_1^2 +c^2\right]^{-s} \label{z_10} \\
  &&Z_{1,\frac12}(a_1,c; s)=\sum_{n_1=0}^\infty \left[a_1^2 (n_1+1/2)^2 +c^2\right]^{-s}.
  \label{z_11/2}
  \eea
  The above  sums can be directly regularized by utilizing two specific  generalized forms of the   Abel-Plana formula \cite{CR1}
\bea
&& \sum_{n=1}^{\infty}( n^2+q^2)^\nu -\int_{0}^\infty du \,(u^2+q^2)^\nu= -\frac{q^{-2s}}{2} -2 \sin \pi \nu \int_{q}^\infty du\, \frac{(u^2-q^2)^\nu}{e^{2\pi u}-1}, \label{AP1} \\
&& \sum_{n=0}^{\infty}\left[\left(n+1/2\right)^2+q^2\right]^\nu -\int_{0}^\infty du \,(u^2+q^2 )^\nu=    2\sin\pi \nu \int_q^\infty du\, \frac{(u^2-q^2)^\nu }{e^{2\pi u}+1},
\label{AP2}
\eea
where  the right-hand sides of the above equations are the regularized forms of the sums. As a result  one can directly obtain
 \bea
&& Z_{1,0}(a_1, c;s)=-\frac{c^{-2s}}{2}+ 2 \sin(\pi s) \int_{c/a_1}^\infty du \frac{ (a_1^2 u^2-c^2)^{-s}}{e^{2\pi u}-1}  \label{rz_10}  \\
  &&Z_{1,\frac12}(a_1,c; s)=-2 \sin(\pi s) \int_{c/a_1} ^\infty du \frac{ (a_1^2 u^2-c^2)^{-s}}{e^{2\pi u}+1}
  \label{rz_11/2}
  \eea
 Subsequent cases  are given by the recurrences \eqref{rz_p0-re} and  \eqref{rz_p1/2-re}.  Some cases needed for the later calculations are given as follow;
  \bea
&& Z_{2,0}(a_1,a_2, c;s)=\frac{c^{-2 s}}{4}-\frac{\sqrt{\pi } c^{1-2 s} }{4 a_1 }\frac{\Gamma \left(s-\frac{1}{2}\right)}{\Gamma (s)}-\sin(\pi s) \int_{c/a_2}^\infty du \frac{ (a_2^2 u^2-c^2)^{-s}}{e^{2\pi u}-1} \nn \\
&&  -\frac{\sqrt{\pi }}{a_1}\frac{\cos (\pi  s) \Gamma \left(s-\frac{1}{2}\right) }{  \Gamma (s)}\int_{c/a_2}^\infty du \frac{ (a_2^2 u^2-c^2)^{\frac12-s}}{e^{2\pi u}-1}  +\frac{\sqrt{\pi } }{2 a_1}\frac{ G_{2,0}\left(a_1,a_2,c;s-\frac{1}{2}\right)}{\Gamma (s)}
 \label{rz_20} 
  \eea
 
   \bea
&& Z_{3,0}(a_1,a_2, a_3, c;s)=-\frac{c^{-2 s}}{8} +\frac{\sqrt{\pi } c^{1-2 s} }{8}\left(\frac{1}{ a_1 }+\frac{1}{ a_2 }\right)\frac{\Gamma \left(s-\frac{1}{2}\right)}{\Gamma (s)} -\frac{\pi  c^{2-2 s} }{8 a_1 a_2}\frac{\Gamma (s-1)}{ \Gamma (s)} \nn \\
&& \hs{3cm}+\frac{\sqrt{\pi }}{2}\left(\frac{1}{a_1}+\frac{1}{a_2}\right)\frac{\cos (\pi  s) \Gamma \left(s-\frac{1}{2}\right) }{  \Gamma (s)}\int_{c/a_3}^\infty du \frac{ (a_3^2 u^2-c^2)^{\frac12-s}}{e^{2\pi u}-1} \nn \\
&& \hs{1cm}+\frac{\sin(\pi s)}{2} \int_{c/a_3}^\infty du \frac{ (a_3^2 u^2-c^2)^{-s}}{e^{2\pi u}-1} -\frac{\pi}{a_1 a_2} \frac{\sin (\pi  s) \Gamma \left(s-1\right) }{  \Gamma (s)}\int_{c/a_3}^\infty du \frac{ (a_3^2 u^2-c^2)^{1-s}}{e^{2\pi u}-1}  \nn \\
&&\hs{2cm}+\frac{\sqrt{\pi } }{2 a_1}\frac{ G_{3,0}\left(a_1,a_2,a_3,c;s-\frac{1}{2}\right)}{\Gamma (s)}  -\frac{\sqrt{\pi }}{4 a_2}\frac{ G_{2,0}\left(a_2,a_3,c;s-\frac{1}{2}\right)}{ \Gamma (s)} \nn \\
&& \hs{9cm}+\frac{\pi}{4 a_1 a_2}\frac{  G_{2,0}\left(a_2,a_3,c;s-1\right)}{ \Gamma (s)}
 \label{rz_30} 
  \eea  
  \bea
&&  Z_{4,0}(a_1,a_2, a_3, a_4, c; s)= -\frac{c^{-2 s}}{16} -\frac{\sqrt{\pi } c^{1-2 s} }{16}\left(\frac{1}{ a_1 }+\frac{1}{ a_2 }+\frac{1}{ a_3 }\right)\frac{\Gamma \left(s-\frac{1}{2}\right)}{\Gamma (s)} \nn \\
&&\hs{2cm}+\frac{\pi  c^{2-2 s} }{16 }\left(\frac{1}{a_1 a_2}+\frac{1}{a_1 a_3}+\frac{1}{a_2 a_3}\right)\frac{\Gamma (s-1)}{ \Gamma (s)}-\frac{\pi^{3/2}  c^{3-2 s} }{16 a_1 a_2 a_3} \frac{\Gamma (s-3/2)}{ \Gamma (s)} \nn \\
&& \hs{2cm}-\frac{\sqrt{\pi }}{4}\left(\frac{1}{a_1}+\frac{1}{a_2}+\frac{1}{a_3}\right)\frac{\cos (\pi  s) \Gamma \left(s-\frac{1}{2}\right) }{  \Gamma (s)}\int_{c/a_4}^\infty du \frac{ (a_4^2 u^2-c^2)^{\frac12-s}}{e^{2\pi u}-1} \nn \\
&& \hs{2cm} +\pi\left(\frac{1}{a_1 a_2}+\frac{1}{a_1 a_3}+\frac{1}{a_2 a_3}\right) \frac{\sin (\pi  s) \Gamma \left(s-1\right) }{  \Gamma (s)}\int_{c/a_3}^\infty du \frac{ (a_3^2 u^2-c^2)^{1-s}}{e^{2\pi u}-1} \nn \\
&&+  \frac{\pi}{a_1 a_2 a_3} \frac{\sin (\pi  s) \Gamma \left(s-1\right) }{  \Gamma (s)}\int_{c/a_4}^\infty du \frac{ (a_4^2 u^2-c^2)^{1-s}}{e^{2\pi u}-1} -\frac{\sin(\pi s)}{4} \int_{c/a_4}^\infty du \frac{ (a_4^2 u^2-c^2)^{-s}}{e^{2\pi u}-1} \nn \\
&& \hs{2cm}+\frac{\sqrt{\pi } }{2 a_1}\frac{ G_{4,0}\left(a_1,a_2,a_3,a_4,c;s-\frac{1}{2}\right)}{\Gamma (s)}-\frac{\sqrt{\pi }}{4 a_2}\frac{ G_{3,0}\left(a_2,a_3,a_4,c;s-\frac{1}{2}\right)}{ \Gamma (s)} \nn \\
&& \hs{2cm}+\frac{\pi}{4 a_1 a_2}\frac{  G_{3,0}\left(a_2,a_3,a_4,c;s-1\right)}{ \Gamma (s)} +\frac{\sqrt{\pi } }{8a_3}\frac{ G_{2,0}\left( a_3,a_4,c;s-\frac{1}{2}\right)}{\Gamma (s)}\nn \\
&&-\frac{\pi}{8}\left(\frac{1}{  a_1 a_3}+\frac{1}{  a_2 a_3}\right)\frac{  G_{2,0}\left(a_3,a_4,c;s-1\right)}{ \Gamma (s)} +\frac{\pi^{3/2}}{8 a_1 a_2 a_3} \frac{  G_{2,0}\left(a_3,a_4,c;s-\frac32\right)}{ \Gamma (s)} 
  \label{rz_40}
  \eea
 
    Note that some  infinite sums  of the form \eqref{z_p} has been studied in \cite{math2} in a rather different manner based on the Riemann and/or the Hurwitz zeta regularization. However, the resulting expressions of the mentioned work are well-defined for $c\neq0$, such that  the $c=0$ case has been considered separately.  However,  it is important for our work to find an exact expression for a  finite ``$c$'', resulting in an appropriate regularized  form for   $c \to 0$, as  here ``$c$'' has the role of the  mass parameter. 
    
  Note that    the summations over the entire integer numbers can also be written  in terms of the above zeta functions. For instance, 
  \bea
  &&\sum_{n_1,n_2,n_3,n_4=-\infty}^\infty \left[a_1^2n_1^2+...+a_4^2n_4^2+c^2\right]^{-s}= c^{-2s}+ 2\Big( Z_{1,0}(a_1,c;s)+Z_{1,0}(a_2,c;s)\nn \\
  && \hs{1cm}+ Z_{1,0}(a_2,c;s)+Z_{1,0}(a_2,c;s) \Big) +4 \Big( Z_{2,0}(a_1, a_2,c;s)+Z_{2,0}(a_1,a_3,c;s)\nn \\
  && \hs{1cm}+ Z_{2,0}(a_1,a_4,c;s)+Z_{2,0}(a_2,a_3,c;s)+Z_{2,0}(a_2,a_4,c;s) +Z_{2,0}(a_3,a_4,c;s)\Big)\nn \\ 
  && \hs{1cm}+8 \Big(Z_{3,0}(a_1,a_2,a_3,c;s)+ Z_{3,0}(a_1,a_2,a_4,c;s)+ Z_{3,0}(a_2,a_3,a_4,c;s)\Big)\nn \\ 
  &&\hs{9cm}+16 Z_{4,0}(a_1,a_2,a_3,a_4,c;s) ,
    \label{z_40-ent}
    \eea
    while 
    \bea
 \sum_{n_1,...,n_p=-\infty}^\infty \left[a_1^2(n_1+1/2)^2+...+a_p^2(n_p+1/2)^2+c^2\right]^{-s}= 2^{p} Z_{p,\frac12}(a_1,a_2,...,a_p,c;s).
     \label{z_p1/2-ent}
  \eea
Some other cases   needed in this works can be obtained in a similar way; 
  \bea
&&\sum_{n_1=1}^\infty\sum_{n_2=0}^\infty \left[a_1^2 n_1 ^2+a_2^2 (n_2+1/2)^2+c^2\right]^{-s}=\sin(\pi s) \int_{c/a_2}^\infty du \frac{ (a_2^2 u^2-c^2)^{-s}}{e^{2\pi u}+1} \nn \\
&& +\frac{\sqrt{\pi }}{a_1}\frac{\cos (\pi  s) \Gamma \left(s-\frac{1}{2}\right) }{  \Gamma (s)}\int_{c/a_2}^\infty du \frac{ (a_2^2 u^2-c^2)^{\frac12-s}}{e^{2\pi u}+1} +\frac{\sqrt{\pi }}{2 a_1 } \frac{G_{2,\frac12}\left(a_1,a_2, c;s-\frac{1}{2}\right)}{ \Gamma (s)}
  \label{rz_201/2}
  \eea
  \bea
&&\sum_{n_1=0}^\infty\sum_{n_2=1}^\infty \left[a_1^2(n_1+1/2)^2+a_2^2 n_2^2+c^2\right]^{-s}=  -\frac{\sqrt{\pi } c^{1-2 s} }{4 a_1}\frac{\Gamma \left(s-\frac{1}{2}\right) }{\Gamma (s)} \nn \\
 && -\frac{\sqrt{\pi }}{a_1}\frac{\cos (\pi  s) \Gamma \left(s-\frac{1}{2}\right) }{  \Gamma (s)}\int_{c/a_2}^\infty du \frac{ (a_2^2 u^2-c^2)^{\frac12-s}}{e^{2\pi u}-1} -\frac{\sqrt{\pi }}{2 a_1 } \frac{G_{2,0}\left(a_1,a_2, c;s-\frac{1}{2}\right)}{ \Gamma (s)}.
  \label{rz_21/20}
  \eea
Note that  for  $a_1\ll a_2\ll  a_3 \ll a_4$  all the $G$ function terms in the above zeta expressions  can be neglected  with a good degree of approximation, see   Eq. \eqref{G}. As a result, as we see in the next sections, by using the above zeta functions one can appropriately regularize the vacuum energy obtaining two different representation   asymptotically suitable for low/high temperatures. 
  Finally to obtain the Casimir energy  one must  also  subtract the contribution terms of the free vacuum, which are given by the first few terms of the heat-kernel expansion  \cite{CR1,CRec8,math6}. As we see in the following sections, through the above approach, these contribution terms would be automatically appeared in the resulting  asymptotic  expressions for Casimir energy at high temperature.

\section{Ideal-metal planes}
 
 \subsection{Two parallel planes}
  For two parallel  planes with separation distance ``$a$'', at temperature ``$T$'',  the total    zero-point energy of  electromagnetic field   is well-known;
  \be
E_0 = \frac1{\beta}\frac{A}{2\pi} \int_{0}^{\infty}  k_\perp dk_\perp \sum_{n=1}^\infty \sum_{l=-\infty}^{\infty} \ln\left[  l^2+ \left(\frac{\beta }{2\pi}\right)^2 \left[k_\perp^2+\left(\frac{n \pi}{a}\right)^2  \right] \right],
\label{e_0-p}
 \ee
 in which, ``$A$'' is the area of each plane, and $k_\perp$ is the total momentum of a  mode propagating parallel to the planes. Rewriting the above equation as a parametric integral, similarly as the previous section, one obtains 
 \bea
 E_0 = -  \frac{\pi A}{\beta^3} \lim_{s\rightarrow 0} \frac{\partial}{\partial s} \frac{\Gamma (s-1)}{\Gamma (s)} \left( \sum_{n=1}^\infty  (\lambda_a^2 n^2)^{1-s}+2\sum_{l,n =1}^\infty (l^2+ \lambda_a^2 n ^2 )^{1-s}\right)
\label{pe_0-p}
 \eea
where we have   made use of a Gaussian integral for $k_\perp$, and  separated  the $l=0$ term.  Now   both the sum terms can be directly regularized in terms of  $Z_{1,0}$ and $Z_{2,0}$.   But  according to the previous section,    if  in Eq. \eqref{rz_20} we choose $a_{1,2}=1,\lambda_a$,  the resulting expression would be  asymptotically suitable     for small temperatures ($\lambda_a\geq 1$), while  choosing $a_{1,2}=\lambda_a,1$   would be appropriate  for   large temperatures ($\lambda_a\leq 1$);
\bea
 E_0  = -  \frac{\pi A}{ \beta^3} \lim_{s\rightarrow 0} \frac{\partial}{\partial s}  \frac{\Gamma (s-1)}{\Gamma (s)}\Bigg(Z_{1,0}(\lambda_a,0;s-1)+2 Z_{2,0}(\lambda_a ,1,0 ;s-1)\Bigg).
\label{pe_0-p}
 \eea
  So  using Eqs. \eqref{rz_10} and \eqref{rz_20} in the above equation, and after some   calculations  one can find the regularized  vacuum energy as
\bea
&&E_0(a,T)=-  \frac{ A}{ \beta^3} \Bigg[  \frac{8\pi^2}{3\lambda_a} \int_0^\infty du\frac{ u^3}{e^{2\pi u}-1} +2\pi^2\lambda_a^2  \int_0^\infty du\frac{ u^2  }{e^{2\pi u}-1}  \nn \\
&& \hs{5cm} + \frac{4\pi}{\lambda _a  }\sum_{n,l=1}^\infty \left(\frac{ l }{n/\lambda _a}\right)^{3/2 } K_{3/2}\left(\frac{2 l n \pi }{\lambda _a} \right)\Bigg]
\label{re_0-p2}
\eea
Now one can see that the first     term of the above equation   is  just equal to  the main term of the corresponding  heat-kernel expansion  \cite{CRec8}, i.e.  the free energy contribution of the black body radiation. Note that   unlike the conventional regularization approaches, here the contribution of the free vacuum has been appeared automatically in the regularized zero-point energy.  By subtracting this term  one finds another exact expression for the Casimir energy, which,  in contrast to the familiar expressions of the literature,  has an asymptotically explicit form for sufficiently large temperatures;
  \bea
  E_\trm{C}(a,T)\approx  - A\frac{   \zeta(3)}{8\pi a^2   } T; \ \ \ \ \  T\gg 1/2a .
   \label{ht-e_c-p}
  \eea
 This is the  familiar high-temperature (classical)  limiting expression of the Casimir energy of the parallel planes.

 \subsection{Rectangular box}
  
 The   electromagnetic free energy in a rectangular ideal-metal box, can be written as
  \bea
&&E_0  = \frac{1}{2\beta}  \sum_{l=-\infty}^{\infty}\Bigg( 2 \sum_{n,m,r=1}^\infty  \ln\left[  l^2+\lambda_a^2 n^2+\lambda_b^2 m^2+ \lambda_c^2 r^2  \right] +\sum_{m,r=1}^\infty  \ln\left[  l^2 +\lambda_b^2 m^2+ \lambda_c^2 r^2   \right]\nn \\
&& \hspace{3cm}+\sum_{n,r=1}^\infty  \ln\left[ l^2+\lambda_a^2 n^2+  \lambda_c^2 r^2 \right]+\sum_{n,m=1}^\infty  \ln\left[ l^2+\lambda_a^2 n^2+\lambda_b^2 m^2  \right]\Bigg), 
\label{e_0-b}
 \eea
  in which  $\lambda_a\equiv \beta/2a, \lambda_b\equiv \beta/2b, \lambda_c\equiv \beta/2c $ where $a,b,c$ are the sidelengths of the box ($a\leq b\leq c$). Now according to  the previous section, to find an asymptotically suitable expression for  large temperatures ($\lambda_c \leq \lambda_b \leq\lambda_a \leq 1$), the above equation can be written  as
 \bea
&&E_0   =    -\frac{1}{2\beta} \lim_{s\rightarrow 0} \frac{\partial}{\partial s}    \Big[2Z_{3,0}(\lambda_c ,\lambda_b , \lambda_a ,0 ;s) +4Z_{4,0}(\lambda_c ,\lambda_b , \lambda_a ,1,0 ;s)\nn \\
&&\hs{4cm}+Z_{2,0}(\lambda_b, \lambda_a,0; s)+ 2 Z_{3,0}(\lambda_b ,\lambda_a ,1,0 ;s)\nn \\
&&\hs{4cm}+Z_{2,0}(\lambda_c, \lambda_a,0; s)+ 2 Z_{3,0}(\lambda_c ,\lambda_a ,1,0 ;s)\nn \\
&&\hs{4cm}+Z_{2,0}(\lambda_c, \lambda_b,0; s)+ 2 Z_{3,0}(\lambda_c ,\lambda_b ,1,0 ;s)\Big]
\label{pe_0-b}
 \eea
 The above equation can be directly regularized by using Eqs. \eqref{rz_20}, \eqref{rz_30} and \eqref{rz_40}. As a result one can see that the first three terms of the heat kernel expansion, i.e. the terms proportional  to the volume, total side-area and total side-length of the box,  would be automatically appeared  in the resulting regularized expression,  see  the forth term of Eq. \eqref{rz_20}, the forth and sixth terms of Eq. \eqref{rz_30},  etc.  Then  for sufficiently  large temperatures we find
 \bea
&&E_{\trm{emC}}(a,b,c,T)\approx - \frac{1}{\beta}\Bigg(\frac{\zeta(3)}{8\pi}\frac{bc}{ a^2 } +\frac{\pi}{ 24} \frac{ c}{b} -\frac{\ln(\beta /2a)}{4}-\frac{\ln(\beta/2b )}{4}\nn \\
&& \hspace{1cm}  +\sum_{n,m=1}^{\infty} \Bigg[ \sqrt{ \frac{b}{a}}\sqrt{\frac{n  }{  m }} K_1\left(2\pi n   m \frac c a\right)  +\frac{b}{a}\frac{n  }{  m } K_1\left(2\pi n   m \frac ca\right)+\frac{c}{a}\frac{n   }{  m  }K_1\left(2 m n \pi \frac ba \right) \Bigg] \nn \\ &&\hs{1cm}+2\frac{c}{a} \sum_{n,m,l=1}^\infty \frac{n   }{  \sqrt{m^2+l^2 (c/b)^2}  }K_1\left(2   n \pi \frac ba \sqrt{m^2+l^2 (c/b)^2} \right) \Bigg) ; \ \  T\gg 1/a
\label{ht-e_c-b}
\eea
  From  the above equation, one can see that the Casimir forces, acting on the opposite sides of the box, can be attractive or repulsive, depending on the values of $a$, $b$ and $c$,  which is generally  in agreement with the  known results \cite{CR1,CR3, CRec1,CRec2,CRec3,CRec7}.    For a cube (a=b=c)   the above equation results in a  limiting Casimir pressure   as
  \bea
  P_\trm{C}(T\gg 1/2a) \approx  \frac{0.14 }{a^3}T ,
 \label{ht-p_c-c}
 \eea
Actually by using \eqref{ht-e_c-b} it can be numerically shown that the above equation gives the largest pressure of the box. Note that at room temperature, the above equation is valid  for  $a\gtrapprox 6\mu \trm{m}$.  Comparing the above equation  with  that of the parallel planes, one can see that  the strength of   Casimir pressure of a cube can be larger,  by a factor of $1.5$,  than that of the two parallel planes.

 \subsection{Rectangular tube}
The  electromagnetic free energy in an ideal-metal  rectangular tube  can be written as
 \bea
&&E_0  = \frac{1}{2\beta} \frac{L}{2\pi} \int_{-\infty}^{\infty}du \sum_{l=-\infty}^{\infty}\Bigg( 2 \sum_{n,m=1}^\infty  \ln\left[ l^2+\lambda_a^2 n^2+\lambda_b^2 m^2+ (\beta/2\pi)^2 u^2  \right]\nn \\
&& +\sum_{n=1}^\infty  \ln\left[  l^2+\lambda_a^2 n^2 + (\beta/2\pi)^2 u^2  \right] +\sum_{m=1}^\infty  \ln\left[  l^2 +\lambda_b^2 m^2+ (\beta/2\pi)^2 u^2 \right]\Bigg) 
\label{e_0-t}
 \eea
 in which $\lambda_a\equiv \beta/2a, \lambda_b\equiv \beta/2b$, where $a,b,L$ are the sidelengths of the tube ($a\leq b\ll L$).  Then similarly as before,  a suitable expression for large temperatures ($  \lambda_b \leq\lambda_a \leq 1$) can be obtained by regularizing the above equation as
  \bea
&&E_0   =   -\frac{L \sqrt{\pi}}{2\beta^2}  \lim_{s\rightarrow 0} \frac{\partial}{\partial s}   \frac{\Gamma(s-1/2)}{\Gamma(s)}   \Big[ Z_{1,0}(\lambda_a , 0;s-1/2)+ Z_{1,0}(\lambda_b , 0;s-1/2) \nn \\
&&\hs{4cm} +2 Z_{2,0}(\lambda_a ,1, 0;s-1/2) +2 Z_{2,0}( \lambda_b ,1, 0;s-1/2)\nn \\
&&\hs{4cm}+2 Z_{2,0}(\lambda_b ,\lambda_a , 0;s-1/2)+ 4 Z_{3,0}( \lambda_b , \lambda_a , 1, 0;s-1/2)\Big].
 \label{pe_0-t}
\eea
  As a result, after some   calculations one can find
 \bea
 E_\trm{C}(a,b,T)\approx -\frac{L}{2\beta} \Bigg(\frac{\pi  }{12 b}+\frac {\zeta(3)}{4\pi }\frac {b}{a^2}+\frac2a\sum_{n,m=1}^\infty \frac{m }{n}  K_1\left(2 m n \pi \frac{ b}{a}\right)\Bigg); \ \ \ T\gg 1/2a.
 \label{ht-e_c-t}
 \eea
 One can see that  the Casimir pressure for the rectangular tube can be  attractive or repulsive depending on the values of the sidelengths ``$a$'' and ``$b$''.  The limiting Casimir pressure on two parallel sidewalls  with the smallest  separation distance, would be obtained as
\bea
 P_{\trm{C},a}  \approx -\frac{V(b/a)}{a^3}T ; \ \ \ \ T\gg 1/2a
 \label{ht-p_c-t}
 \eea
 in which,
  \bea
 V(x) \equiv\frac{\pi}{24 x}+\frac{ \zeta(3)}{4\pi } - \sum_{n,m=1}^\infty\left[  \pi m^2 K_0(2 m n \pi  x)+  \pi m^2 K_2(2 m n \pi  x)   -\frac{m   K_1(2 m n \pi  x)}{  n x}\right].
\label{V}
\eea
At room temperature,  Eq. \eqref{ht-p_c-t} is valid  for   $a,b\gtrapprox 6\mu \trm{m}$.   By numerical computations one can find  $ V(x )\leq V(1) \approx 0.22$.  So  for  sufficiently  large temperatures, the pressure of the tube can be up to  twice that of the parallel planes. 

 Note that  the high-temperature limiting expressions \eqref{ht-e_c-b} and   \eqref{ht-e_c-t} for the Casimir free energies of the box and tube, respectively, have not been given in other works.  The familiar equations  for the Casimir energy of the mentioned configurations, are asymptotically suitable for small temperatures, see e.g. \cite{CR1,CR2,CRec8}. These equations can also be obtained consistently through the approach of  the section 2.

  \section{Non-Euclidean spaces}
  In this section we  investigate the Casimir energy in spaces with nontrivial topology and curvature.  We restrict ourselves on two  familiar cases i.e. $3$-torus and $3$-sphere, which are of specific importance in some cosmological models.

  \subsection{ $3$-torus}
  As we know, on a compact manifold  with the topology of a torus, a scalar/spinor  field satisfies a periodic  identification condition in all coordinates, that is, for a field $\phi$ on a $d$-torus one can write
  \bea
  \phi(x_i+ a_i) = \phi(x_i) , \ \  \partial_{i}\phi|_{x_i+a_i}=\partial_{i}\phi|_{x_i}; \ \ \ i=1,2,..., d
  \nn
  \eea
  in which $x_i$'s are the space coordinates, $a_i$'s are the identification parameters of the torus, and $\partial_i \equiv \frac{\partial}{\partial x_i}$.  As a result  the  free energy of a scalar  field on a $3$-torus can be written as 
  \bea
  E_0 =\frac{1}{2\beta}\sum_{l,m,n,r=-\infty}^\infty \ln\left[l^2+\lambda_a^2 m^2+\lambda_b^2 n^2+\lambda_c^2 r^2+ \lambda_\mu^2\right],   \label{e_0-scal-tor}
\eea
in which $\lambda_{a,b,c}$ are given as before, and $\lambda_\mu=\beta\mu/2\pi$.  For a $4$-spinor the above equation is  multiplied by  ``$-4$''. Then similarly as before, after some calculations     using Eq. \eqref{z_40-ent}, for sufficiently small   temperatures one obtains
  \bea
 && E_\trm{C}(a,b,c,T) \approx -\frac{1}{2\beta} \Bigg(\frac{4 \pi ^2 \lambda _{\mu }^3}{3 \lambda_a\lambda _b \lambda _c}-\frac{\pi  \lambda _{\mu }^2}{\lambda_a\lambda _c}\big(1- 2  \ln \left(\lambda _{\mu }\right) \big)-\frac{2\pi  \lambda _{\mu }}{ \lambda _c} \nn \\
  &&\hs{1cm}-\frac{16 \pi ^2}{3 \lambda _a \lambda _b \lambda _c}\int_{\lambda _{\mu } }^{\infty }  du \frac{ \big[u^2- \lambda _{\mu } ^2\big]^{3/2}}{\exp (2 \pi  u)-1 }+\frac{16 \pi ^2 \lambda _a^3}{3 \lambda _b \lambda _c}\int_{\lambda _{\mu }/\lambda _a}^{\infty }  du \frac{ \big[u^2-(\lambda _{\mu }/\lambda _a)^2\big]^{3/2}}{\exp (2 \pi  u)-1 }\nn \\
  &&\hs{1cm}+\frac{4 \pi ^2\lambda_b^2}{\lambda _c}\int_{\lambda _{\mu }/\lambda _b}^{\infty }  du \frac{  u^2-(\lambda _{\mu }/\lambda _b)^2 }{\exp (2 \pi  u)-1 }+8\pi \lambda_c \int_{\lambda _{\mu }/\lambda _c}^{\infty }  du \frac{ \big[u^2-(\lambda _{\mu }/\lambda _c)^2\big]^{1/2}}{\exp (2 \pi  u)-1 }\nn \\
  &&+\frac{  2\pi}{\lambda_c}\Big( G_{2,0}(\lambda_c,\lambda_a ,\lambda_\mu;-1)+G_{2,0}(\lambda_c,\lambda_b, \lambda_\mu;-1)+2G_{3,0}(\lambda_c,\lambda_b,\lambda_a ,\lambda_\mu;-1)\Big)\nn \\
  && \hs{5cm}+\frac{ 2\pi^{3/2}}{\lambda_b \lambda_c} G_{2,0}(\lambda_b,\lambda_a,\lambda_\mu; -3/2)\Bigg); \ \ \lambda_{a,b,c}\gg 1
  \label{lt-e_c-scal-tor}
    \eea
    while for large temperatures we  find
  \bea
&& E_\trm{C}(a,b,c,T) \approx -\frac{1}{2\beta} \Bigg( \frac{4 \pi ^2 \lambda_a^2}{\lambda _b \lambda _c}\int_{\lambda _{\mu }/\lambda _a}^{\infty }  du \frac{  u^2-(\lambda _{\mu }/\lambda _a)^2 }{\exp (2 \pi  u)-1 }\nn \\
&&+\frac{8 \pi \lambda_b}{\lambda _c}\int_{\lambda _{\mu }/\lambda _b}^{\infty }  du \frac{ \big[u^2-(\lambda _{\mu }/\lambda _b)^2\big]^{1/2}}{\exp (2 \pi  u)-1 }+4\pi\left( \int_{\lambda _{\mu }/\lambda _c}^{\infty } - \int_{\lambda _\mu }^{\infty } \right)\frac{ du}{\exp (2 \pi  u)-1 }\nn \\
&& +\frac{ 2\pi^{1/2}}{\lambda_c} \Big(G_{2,0}(\lambda_c,\lambda_a,\lambda_\mu; -1/2)+G_{2,0}(\lambda_c,\lambda_b,\lambda_\mu;-1/2)+2G_{3,0}(\lambda_c,\lambda_b,\lambda_a,\lambda_\mu; -1/2)\Big)\nn \\
&&\hs{7cm}+\frac{ 2\pi}{\lambda_b \lambda_c}  G_{2,0}(\lambda_b,\lambda_a,\lambda_\mu;-1) \Bigg) ; \  \lambda_{a,b,c}\ll 1.
\label{ht-e_c-scal-tor}
 \eea  
 Here we have subtracted the contribution of the free vacuum, that is, the vacuum of a $3$-torus with $a,b,c\to \infty$,
 \bea
  && E_0^\trm{f}(a,b,c,T) = -\frac{1}{2\beta} \Bigg(\frac{-4 \pi ^2 \lambda _{\mu }^3}{3 \lambda _a\lambda _b \lambda _c}+\frac{\pi  \lambda _{\mu }^2}{\lambda _a\lambda _c}\big(1- 2  \ln \left(\lambda _{\mu }\right) \big)+\frac{2\pi  \lambda _{\mu }}{ \lambda _c} \nn \\
  &&\hs{7cm}+\frac{16 \pi ^2}{3 \lambda _a \lambda _b \lambda _c}\int_{\lambda _{\mu } }^{\infty }  du \frac{ \big[u^2- \lambda _{\mu } ^2\big]^{3/2}}{\exp (2 \pi  u)-1 }\Bigg).
 \label{e_f-scal-tor}
 \eea
 
  In the case with sufficiently large values of $b/a$, $c/a$ and $c/b$,   all the remaining  G-function terms can also be neglected. As a result  the limiting Casimir energies \eqref{lt-e_c-scal-tor} and \eqref{ht-e_c-scal-tor} for    small masses can be approximated as
  \bea
  E_\trm{C}(a,b,c,T) \approx -\frac{\pi ^2 b c}{180 a^3}-\frac{c \zeta (3)}{4 \pi  b^2}-\frac{\pi }{12 c}+c \mu  T+\frac{4\pi ^2 a b c}{45}  T^4 ; \ \ \mu\ll   1/c \gg T
  \label{lm-lt-e_c-scal-tor}
  \eea
      \bea
  E_\trm{C}(a,b,c,T) \approx T\left[-\frac{b c  \zeta (3)}{2 \pi  a^2}-\frac{\pi  c  }{6 b}+  \ln (   c T)+ \left(\frac{2 }{\pi }-1\right)c \mu\right] ; \ \ \mu\ll    1/a \ll T, 
  \label{lm-ht-e_c-scal-tor}
  \eea
while for   large masses we find
\bea
 && E_\trm{C}(a,b,c,T) \approx-\frac{b c }{2 }\left(\frac{\mu }{\pi   a}\right)^{3/2}  e^{-2 a \mu }-\frac{c \mu }{2 \pi  b} e^{-2 b \mu }-\sqrt{\frac{\mu}{4\pi c}} e^{-2 c \mu } \nn \\
&&  \hs{7cm}+\frac{b c \mu ^3}{3 \pi }-\frac{4 a b c \mu ^3 }{3 \pi }T ; \ \ \mu\gg    1/c \gg T
  \label{hm-lt-e_c-scal-tor}
  \eea
    \bea
  E_\trm{C}(a,b,c,T) \approx -T\left[ c \sqrt{\frac{\mu}{\pi b}} e^{-2 b \mu }+ e^{-2 c \mu }+\frac{b c \mu   }{\pi  a}e^{-2 a \mu } \right] ; \ \ \mu\gg   1/a \ll T, 
  \label{hm-ht-e_c-scal-tor}
  \eea
Eq. \eqref{lm-lt-e_c-scal-tor} is in agreement with the  results of \cite{C-tor1} for the zero-temperature massless case.

 From Eqs. \eqref{lm-lt-e_c-scal-tor} and \eqref{hm-lt-e_c-scal-tor} the  limiting scalar Casimir entropies, $ S_\trm{C}  \equiv-(\partial/\partial T) E_\trm{C}$, on the $3$-torus  are obtained as
  \bea
 &&S_\trm{C} \approx -c\mu -\frac{16\pi^2 a b c}{45} T^3 ; \hs{3.5cm}\mu\ll    1/c \gg T, \nn \\
&& S_\trm{C} \approx -\ln(c T)  ; \hs{5.2cm} \mu\ll    1/a \ll T .
\label{lim-s_c-scal-tor}
 \eea
As is seen, for $\mu\neq 0$  the  scalar Casimir entropy of the torus  takes nonzero values at zero temperature, violating the Nernst heat theorem. One can see that the violating term, i.e. the first term of the first line of the above equation, actually comes from   the thermal contribution of the free vacuum \eqref{e_f-scal-tor}.  Also the entropy takes negative values as well for low as for high temperatures.  Such a negative Casimir entropy appears also in some familiar configuration (in the Euclidean $3$-space), such as the two parallel planes at sufficiently small temperatures,  see e.g. \cite{negS1,negS2,negS3,negS4,negS5}. In Ref. \cite{negS5} the  negative Casimir entropy for  the   parallel   planes case has been justified by assigning a  positive contribution to the conductor walls, such that the total  entropy would be positive. However for a compact manifold such as the $3$-torus the Casimir effect  is not induced  by any  physical boundary walls.  But  the negative entropy  of the torus arises actually as a result of  subtracting   the large positive contribution of the free vacuum, from the total scalar entropy of the $3$-torus. That is, the  scalar entropy of the torus is lower than that of  the free vacuum.    As a result,  we speculate that here the negative entropy can be interpreted as   an instability of the   scalar vacuum state on the torus.  Note that for a spinor field, the above Casimir entropies should be multiplied just by ``-4'',   in agreement  with  the results of \cite{C-tor2} for the zero-temperature massless case. Hence  the spinor vacuum entropy of the $3$-torus is positive  for sufficiently small  as well as large temperatures.

 \subsection{ $3$-sphere}
 The  $00$-component of the  energy-momentum tensor for the vacuum state of the scalar and spinor field on a $3$-sphere are given as \cite{C-sph1,C-sph2}
 \bea
&& \langle0|T^0_0|0\rangle=\frac{1}{4\pi^2 a^3}\sum_{n=1}^\infty n^2 \omega_n; \ \ \omega_n^2=\mu^2+\frac{n^2}{a^2} \label{0t-e_0-scal-sph} \\
&& \langle0|T^0_0|0\rangle=-\frac{1}{\pi^2 a^3}\sum_{n=1}^\infty\left[\left(n+\frac12\right)^2-\frac14\right]\omega_n; \ \ \   \omega_n^2=\mu^2+\frac{(n+1/2)^2}{a^2} \label{0t-e_0-spin-sph}
 \eea
respectively, in which ``$a$'' is the sphere radius, and $\mu$ is the mass parameter.   Therefore the  free energies of the scalar and spinor field can be given by
 \bea
&& E_0 =\frac{1}{2\beta }\sum_{l=-\infty}^\infty \sum_{n=1}^\infty n^2 \ln[l^2+ \lambda_a^2 n^2 +\lambda_\mu^2]  \label{e_0-scal-sph} \\
&& E_0 =-\frac{2}{ \beta }\sum_{l=-\infty}^\infty \sum_{n=1}^\infty\left[\left(n+\frac12\right)^2-\frac14\right]\ln\left[l^2+ \lambda_a^2 \left(n+\frac12\right)^2 +\lambda_\mu^2 \right]  \label{e_0-spin-sph}
 \eea
respectively, in which $\lambda_a\equiv \beta/2\pi a $, and $\lambda_\mu\equiv \beta \mu/2\pi$ as before. Then, similarly as carried out in the section 3,  by rewriting the above equations as parametric integrals and using some relations such as
\bea
n^2\sum_n e^{-t \lambda_a^2 n^2}= -\frac{1}{2 t \lambda_a} \frac{\partial}{\partial \lambda_a} \sum_n e^{-t \lambda_a^2 n^2}\nn 
\eea
  one can rewrite Eqs. \eqref{e_0-scal-sph} and \eqref{e_0-spin-sph}   as
\bea
  E_0(a,T)=\frac{1}{4 \lambda_a\beta }  \frac{\partial}{\partial \lambda_a} \lim_{s\to 0} \frac{\partial}{\partial s}\frac{\Gamma(s-1)}{\Gamma(s)}\sum_{l=-\infty}^\infty \sum_{n=1}^\infty [l^2+ \lambda_a^2 n^2 +\lambda_\mu^2]^{-(s-1)} 
   \label{pe_0-scal-sph}
\eea
and
\bea
&&  E_0(a,T)=-\frac{2}{   \beta }   \lim_{s\to 0} \frac{\partial}{\partial s }\Bigg(\frac{1}{2\lambda_a}\frac{\partial}{\partial \lambda_a}\frac{\Gamma(s-1)}{\Gamma(s)}\sum_{l=-\infty}^\infty \sum_{n=1}^\infty [l^2+ \lambda_a^2 (n+1/2)^2 +\lambda_\mu^2]^{-(s-1)} \nn \\ 
&& \hs{6cm}+\frac14 \sum_{l=-\infty}^\infty \sum_{n=1}^\infty [l^2+ \lambda_a^2 (n+1/2)^2 +\lambda_\mu^2]^{-s}\Bigg) \label{pe_0-spin-sph}
\eea
respectively. Now after separating the $l=0$ term, the above equations can be regularized   in terms of $Z_{2,0}$, $Z_{1,0}$ and $Z_{1,\frac12}$  giving two different representations as well for the scalar as for the spinor  Casimir energy. \\

For the scalar case by using Eqs. \eqref{rz_10} and \eqref{rz_20}  for Eq. \eqref{pe_0-scal-sph} after some calculations, the  limiting Casimir free energies are obtained as 
\bea
&& E_\trm{C}(a,T) \approx  \frac{\pi}{2\beta \lambda_a} \frac{\partial}{\partial \lambda_a} \Bigg( \frac{  \lambda _{\mu }^3}{3 \lambda _a}-\frac{4}{3 \lambda _a}\int_{\lambda _{\mu }}^{\infty } \frac{  \left[u^2-\lambda_\mu^2\right]^{3/2}}{ \exp (2 \pi  u)-1 } \nn \\
&&\hs{5cm} + \frac{4 \lambda _a^3}{3}  \int_{ \lambda _\mu /\lambda _a}^{\infty }  du \frac{\left[u^2-(\lambda _\mu /\lambda _a)\right]^{3/2}}{\exp (2 \pi  u)-1}\Bigg); \ \  \lambda_a\gg 1,
 \label{lt-e_c-scal-sph}
\eea
 \bea
  E_\trm{C}(a,T) \approx \frac{\pi}{2\beta \lambda_a} \frac{\partial}{\partial \lambda_a} \Bigg(   \lambda _a^2 \int_{\lambda _\mu/  \lambda _a}^{\infty } du \frac{  u^2 -(\lambda _\mu/\lambda_a) ^2 }{\exp (2 \pi  u)-1}\Bigg); \ \  \lambda_a\ll 1.
 \label{ht-e_c-scal-sph}
\eea
Here  the energy contribution of the free scalar vacuum i.e. the vacuum energy contribution  of a  $3$-sphere with infinitely large radius $a\to \infty$, 
\bea
  E_0^\trm{f}(a,T) = - \frac{1}{4\beta \lambda_a} \frac{\partial}{\partial \lambda_a} \Bigg( \frac{2 \pi  \lambda _{\mu }^3}{3 \lambda _a}-\frac{8\pi}{3 \lambda _a}\int_{\lambda _{\mu }}^{\infty } \frac{  \left[u^2-\lambda_\mu^2\right]^{3/2}}{ \exp (2 \pi  u)-1 }\Bigg)
\label{e_f-scal-sph}
\eea
has been subtracted. Then for sufficiently small masses the above limiting Casimir energies take the forms
\bea
 E_\trm{C}(a,T) \approx \frac{1}{240 a}+\frac{\pi ^4 a^3}{45}  T^4-\left(\frac{1}{9}  +\frac{\pi}{6} \right)   a^3 \mu ^3 T; \ \ \mu\ll   1/a \gg T
 \label{lm-lt-e_c-scal-sph}
 \eea
 \bea
 E_\trm{C}(a,T) \approx\frac{\zeta (3)}{4 \pi ^2   } T; \ \ \mu\ll   1/a \ll T,
 \label{lm-ht-e_c-scal-sph}
 \eea
 while for sufficiently large masses we find
 \bea
 E_\trm{C}(a,T) \approx \frac{a^{3/2} \mu^{5/2}}{4 \pi  }e^{-2 \pi  a \mu }-\frac{ \pi  a^3 \mu ^3}{6} T ; \ \ \mu\gg  1/a \gg T
 \label{hm-lt-e_c-scal-sph}
 \eea
 \bea
 E_\trm{C}(a,T) \approx\frac{a^2 \mu ^2 T}{2} e^{-2 \pi  a \mu } ; \ \ \mu\gg   1/a \ll T,
 \label{hm-ht-e_c-scal-sph}
 \eea
\\

For the spinor field, from Eq. \eqref{pe_0-spin-sph} and after some similar calculation using Eqs  \eqref{rz_201/2}  and \eqref{rz_21/20}, we find
\bea
&& E_\trm{C}(a,T) \approx   -\frac{2}{ \beta} \Bigg(\frac{1}{2\lambda_a} \frac{\partial}{\partial \lambda_a}\Bigg[\frac{2 \pi  \lambda _{\mu }^3}{3 \lambda _a}-\frac{8\pi}{3\lambda_a}\int_{\lambda _{\mu }}^{\infty } du \frac{ ( u^2-\lambda_\mu^2)^{3/2}}{ \exp (2 \pi  u)+1 }\nn \\
&&-\frac{8\pi \lambda_a^3}{3}\int_{ \lambda _\mu /\lambda _a}^{\infty }du  \frac{ \big[u^2-(\lambda _\mu /\lambda _a)^2]^{3/2}  }{ \exp (2 \pi  u)+1 } \Bigg]-\pi \lambda_a \int_{ \lambda _ \mu /\lambda _a }^{\infty } du \frac{ \big[u^2-( \lambda _ \mu /\lambda _a )^2\big]^{1/2} }{\exp (2 \pi  u)+1} \nn \\
&&\hs{4cm} -\frac{\pi  \lambda _{\mu }}{4 \lambda _a} -\frac{\pi }{\lambda _a}\int_{\lambda _{\mu }}^{\infty } du\frac{ [ u^2-\lambda_\mu^2]^{1/2}}{ \exp (2 \pi  u)+1}  \Bigg); \ \ \lambda_a\gg 1,
 \label{lt-e_c-spin-sph}
\eea
\bea
&& E_\trm{C}(a,T) \approx   \frac{2}{ \beta} \Bigg(\frac{1}{2\lambda_a} \frac{\partial}{\partial \lambda_a}\Bigg[2 \pi  \lambda _a^2 \int_{\lambda _\mu/\lambda _a}^{\infty } du  \frac{  u^2-(\lambda _{\mu }/ \lambda _a)^2 }{\exp (2 \pi  u)+1}\Bigg]\nn \\
 &&\hs{4cm}+\frac{\pi }{2 }\int_{ \lambda _\mu/\lambda _a }^{\infty }  \frac{du }{ \exp (2 \pi  u)+1 } \Bigg); \ \ \lambda_a \ll 1.
 \label{ht-e_c-spin-sph}
\eea
 Here the contribution of the free spinor vacuum energy of the $3$-sphere is given as
 \bea
&& E_0^\trm{f}(a,T) =   -\frac{2}{ \beta} \Bigg(\frac{1}{2\lambda_a} \frac{\partial}{\partial \lambda_a}\Bigg[-\frac{2 \pi  \lambda _{\mu }^3}{3 \lambda _a}+\frac{8\pi}{3\lambda_a}\int_{\lambda _{\mu }}^{\infty } du \frac{ ( u^2-\lambda_\mu^2)^{3/2}}{ \exp (2 \pi  u)+1 }\Bigg]\nn \\
&& \hs{4cm} +\frac{\pi  \lambda _{\mu }}{4 \lambda _a} +\frac{\pi }{\lambda _a}\int_{\lambda _{\mu }}^{\infty } du\frac{ [ u^2-\lambda_\mu^2]^{1/2}}{ \exp (2 \pi  u)+1}  \Bigg).
 \label{e_f-spin-sph}
\eea
 Then for sufficiently small masses, the spinor Casimir energies \eqref{lt-e_c-spin-sph} and \eqref{ht-e_c-spin-sph} take the forms
 \bea
 E_\trm{C}(a,T) \approx   \frac{17}{480 a}+\frac{ \pi ^2 a}{12} T^2  +\frac{\pi  a \mu}{2}   T; \ \ \mu\ll  1/a \gg T.
 \label{lm-lt-e_c-spin-sph}
\eea
\bea
 E_\trm{C}(a,T) \approx \frac{ 3 \zeta (3)+\pi ^2 \ln (4) }{4 \pi ^2} T; \ \ \mu\ll  1/a \ll T,
 \label{lm-ht-e_c-spin-sph}
\eea
while for sufficiently large masses we find
\bea
 E_\trm{C}(a,T) \approx  \frac{a^{3/2} \mu ^{5/2} }{\pi }e^{-2 \pi  a \mu }+\frac{2}{3} \pi  a^3 \mu ^3 T ; \ \ \mu\gg   1/a \gg T.
 \label{hm-lt-e_c-spin-sph}
\eea
\bea
 E_\trm{C}(a,T) \approx 2 a^2 \mu ^2  e^{-2 \pi  a \mu }T; \ \ \mu\gg  1/a \ll T,
 \label{hm-ht-e_c-spin-sph}
\eea
  Eqs. \eqref{lm-lt-e_c-scal-sph}, \eqref{hm-lt-e_c-scal-sph}, \eqref{lm-lt-e_c-spin-sph} and \eqref{hm-lt-e_c-spin-sph}  are in agreement with the zero-temperature results of \cite{C-sph1,C-sph2,C-sph3,C-sph4,C-sph5}. However the temperature corrections obtained here  for the Casimir energy of the sphere as well as torus, have not been given  in other works.  \\
 
  From Eqs. \eqref{lm-lt-e_c-scal-sph} and \eqref{lm-ht-e_c-scal-sph} the limiting scalar   Casimir entropies on the $3$-sphere  are given as
  \bea
 &&S_\trm{C} \approx \left(\frac{1}{9}  +\frac{\pi}{6} \right)   a^3 \mu ^3 - \frac{4\pi ^4 a^3}{45}  T^3 ; \hs{1cm}\mu\ll   1/a \gg T\nn \\
&& S_\trm{C} \approx -\frac{\zeta (3)}{4 \pi ^2   } ; \hs{4.5cm} \mu\ll   1/a \ll T .
\label{lim-s_c-scal-tor}
 \eea 
 while for the spinor case using  Eqs. \eqref{lm-lt-e_c-spin-sph} and \eqref{lm-ht-e_c-spin-sph}  we find
  \bea
 &&S_\trm{C} \approx -\frac{\pi  a \mu}{2}- \frac{ \pi ^2 a}{6} T  ; \hs{1.5cm}\mu\ll  1/a \gg T\nn \\
&& S_\trm{C} \approx -\frac{ 3 \zeta (3)+\pi ^2 \ln (4) }{4 \pi ^2} ; \hs{1cm} \mu\ll    1/a \ll T.
\label{lim-s_c-scal-tor}
 \eea
 Again, as is seen, a nonzero mass term violates the third law of thermodynamics, and yet the scalar as well as  the spinor entropy of the $3$-sphere is negative, which as before, results from subtracting the contribution of the free vacuum. As before, we interpret this negative entropy   as  an instability  of the finite-temperature vacuum state  of   scalar  as well as spinor fields on the $3$-sphere. Note that  again  the thermal contribution of the free vacuum is responsible for  the terms violating  the basic  laws of thermodynamics.  \\

  \section{Concluding remarks}
 In this work we have introduced a new useful    approach to find   asymptotically suitable expressions for the Casimir free energy for sufficiently large temperatures. This approach is based on a rather different regularization technique for an inhomogeneous Epstein-like zeta function.  This approach works well for many familiar configurations in Euclidean as well as non-Euclidean spaces.   The resulting expressions for the Casimir energy contain the classical terms as well as the  first few terms of the corresponding heat-kernel expansion, as expected. The resulting asymptotic expressions, specially for the classical part, might not be simply obtained through other conventional approaches. By utilizing this technique, we have provided some new numerical results specifically for  the    Casimir pressure inside   rectangular ideal-metal   cavities.  We have shown that at sufficiently large temperature, the Casimir pressure in a rectangular tube and a rectangular  box can be larger  by factors of $2$ and $1.5$,  respectively, than that of the two parallel ideal-metal  planes. Note that for sufficiently small temperatures the Casimir pressures of the box and tube are close to that of the parallel planes.    We have also applied this technique   for calculating the Casimir  free energy as well on a $3$-torus  as on a $3$-sphere.  We have shown that a nonzero mass term for both scalar and  spinor fields as well on the torus  as on the sphere, violates the third law of thermodynamics.  We have   obtained    negative values for the scalar Casimir entropy   on the torus,  and for both scalar and  spinor Casimir entropies  on the torus  as well as   on the sphere.   Actually, these negative entropies arise when one subtracts the contribution of the free vacuum.  This means that the vacuum entropy of the system is lower than that of the free vacuum.   Thermodynamically we speculate that this negative Casimir entropy can be interpreted as an instability of the finite-temperature vacuum state. 

 \section*{Acknowledgment}
 We thank G. Jafari for his valuable comments, and S. Qolibikloo for his helps during
the numerical computations

\end{document}